# Twist Defect in Chiral Photonic Structures

Victor I. Kopp and Azriel Z. Genack

*Chiral Photonics, Inc., Clifton, New Jersey 07012*

*Department of Physics, Queens College of CUNY, Flushing, New York 11367*

We demonstrate that twisting one part of a chiral photonic structure about its helical axis produces a single circularly polarized localized mode that gives rise to an anomalous crossover in propagation. Up to a crossover thickness, this defect results in a peak in transmission and exponential scaling of the linewidth for a circularly polarized wave with the same handedness as structure. Above the crossover, however, the linewidth saturates and the defect mode can be excited only by the oppositely polarized wave, resulting in a peak in reflection instead of transmission.

Localized modes produced by regular defects in periodic structures can serve as low threshold lasers [1], low loss waveguides, and narrow band filters in an expanding array of photonic devices [2, 3]. When defects are randomly positioned, localized states are produced within the photonic band gap (PBG) by multiple scattering [4, 5]. The photon dwell time at resonance with a defect state generally increases exponentially with the structure thickness $L$. This suppresses the lasing threshold in an active structure without the use of external feedback. In 3D PBGs, thresholdless lasing has been predicted when the band gap exceeds the emission bandwidth since emission can then couple only to the single defect mode [1]. Exceptionally low lasing thresholds have been achieved in vertical cavity surface emitting lasers (VCSELs), in which an amplifying defect layer is sandwiched between stacks of layers with periodically alternating refractive index [6]. A rich variety of physical effects have been observed when point, line and planar defects are introduced into periodic dielectric structures by the modifying refractive index or thickness of *isotropic* material. We consider here the unusual photonic properties of a defect created by twisting one part of an *anisotropic* structure relative to the rest of the sample. This "chiral twist" provides an additional degree of freedom for the design of photonic structures.

Periodic chiral media appear in nature and can be synthesized as self-organized cholesteric liquid crystals (CLCs) [7] or fabricated by a variety of techniques, for example, by glancing-angle deposition on a rotating substrate [8]. In CLCs, the director, which is the average orientation of molecules in a plane, rotates with pitch $P$ as shown in Fig. 1a. A reflection band forms for circularly polarized light with the same handedness as the structure itself. Light with opposite circular polarization is transmitted through the structure without attenuation. The edges of the band are defined by two sharp optical modes peaked at a wavelength inside the medium equal to $P$. These correspond to circularly polarized standing waves, in which the electric field in the bulk of the sample is aligned along either the ordinary or the extraordinary molecular axes [9]. It was in dye-doped CLCs that suppression of the density of states in a stop band and lasing in long-lived modes at the band edge were first demonstrated [10]. In analogy with isotropic periodic structures, a defect can be produced in a helical structure by adding an isotropic layer in the middle of a CLC [11, 12]. The reflection and transmission of radiation with opposite senses of circular polarization were shown to be equal at resonance [12]. In addition, a "chiral twist" defect created by rotating one part of the sample about its helical axis without separating the two parts was proposed (Fig. 1) [11, 13]. Recently, such a spacerless defect has been created in sculptured thin films composed of multiple helical columns [13], which have optical properties similar to anisotropic chiral media [14]. The peak in transmission was observed for circularly polarized light with the same handedness as the structure and related to the phase shift between reflected waves from the two parts of the structure considered separately [13]. . Here we demonstrate that new phenomena emerge when we consider the optical interaction with the structure as a whole.

In this Letter, we use scattering and transfer matrix calculations to demonstrate that the "chiral twist" defect results in a new type of localized state with a peak in reflection instead of transmission beyond a crossover length $L_{co}$. For specificity, we restrict our discussion to samples with a chiral twist of $90°$ in the center of the sample, which creates a photonic defect at a frequency in the center of the stop band. More generally, varying the chiral twist angle from $0^0$ to $180^0$ tunes the defect frequency from the low to the high frequency band edge. The chiral twist defect introduces a single localized mode into the band gap with a polarization with the same handedness as the structure. In contrast, an additional spacing introduced into isotropic periodic structures creates a degenerate pair of modes. As a result, the polarization of the chiral twist defect mode is independent of the polarization of the exciting mode, in contrast to the polarization of the wave inside a 1D isotropic system, which matches that of the exciting radiation. Below $L_{co}$, the localized mode is excited only by a wave with the same handedness as the structure and exhibits an exponential spatial distribution of the energy density and a peak in transmission at the defect frequency (Fig. 2a). Above the crossover, however, the defect mode can be excited only by the oppositely polarized wave and a resonant peak appears in reflection (Fig. 2b). Though the polarization is transformed within the sample, the polarization of the transmitted and reflected waves is the same as that of the incident wave.

We use scattering [9] and transfer matrix [15] approaches to calculate the energy density inside a right-handed CLC sample and to decompose that density into components of different circular polarization and propagation direction. Identical results are obtained for short structures; however, the transfer matrix approach becomes unstable for thick structures for which we use the scattering matrix method exclusively. The indices of refraction of the sample are taken to be $n_o$ = 1.52 and $n_e$ = 1.58, which are typical values for CLCs. The sample is not absorbing and is nearly index matched to its surroundings, which has a background index of refraction $n_b = n \equiv (n_o + n_e)/2$ = 1.55. The sample pitch is $P$ = 400 nm, placing the center of the band gap at a wavelength of $\lambda_c$ = nP = 620 nm.

With a chiral twist of $90°$ at its physical center, a peak appears in transmission for RCP radiation at $\lambda_c$ for $L < L_{co}$ (Fig. 2a). The crossover in propagation is found in simulations of transmission of right circularly polarized (RCP) radiation and reflection of left circularly polarized (LCP) radiation versus $L$ at the frequency of the localized state (Fig. 3). Figure 2b shows the disappearance of the peak in transmission for RCP light and the growth of a peak in reflection for LCP light beyond $L_{co}$. The variation with thickness of transmittance and reflectance at the resonant peak is displayed in Fig. 3. The crossover thickness, defined as the thickness at which resonant transmission and reflection equal 0.5, is seen in the figure to be 41.25 $P$. The sum of transmission of RCP light and reflection of LCP light is shown as the dotted line in Fig. 3 and seen to equal unity. The reason for this will be explained below in terms of the coupling of the wave to a localized defect mode. In addition, resonant transmission for the right and left circular polarized wave and similarly the reflection for both polarizations are equal for any thickness.

The crossover behavior at the chiral twist defect resonance reflects the anomalous energy distribution inside the sample (Fig. 4). The energy densities of the forward propagating wave inside the structure for incident RCP and LCP radiation are shown in Fig. 4a for samples with $L/P$ = 10, 20, 40, 80 and 160. For all samples, the intensity reaches a local maximum at the defect site in the middle of the sample. For $L \ll L_{co}$ the intensity falls exponentially from the physical center of the sample for the incident RCP wave, as would occur for a defect in an isotropic medium, while the intensity for incident LCP light is constant throughout the sample. Thus only the incident RCP wave is effectively coupled to the localized state in this case. This coupling is the cause of the essentially complete transmission of the RCP wave. Since the intensity for the incident RCP wave at resonance scales exponentially, the photon dwell time within the sample does as well. This leads to an exponential rise in the inverse linewidth seen in Fig. 5. For $L \geqslant L_{co}$, the intensity inside the sample for an incident RCP wave is everywhere smaller than the intensity of the incident wave. This indicates that the incident RCP wave and the localized mode become decoupled beyond the crossover. In contrast, the intensity for the incident LCP wave increases towards the defect site by more than three orders of magnitude. This reflects the strong coupling of this wave to the localized mode, which causes strong reflection and weak transmission of the LCP wave. Beyond $L_{co}$, the peak intensity inside the structure as well as the mode linewidth, which is the same in reflection and transmission, saturates as seen in Fig. 5.

To further analyze the nature of reflection of LCP and transmission of RCP radiation at the defect wavelength, we introduce weak absorption into the chiral structure. The results are shown in Fig. 6 for $L = L_{co}$ for an absorption coefficient of 3 cm$^{-1}$, which is sufficiently small that it only affects long-lived photonic modes. It can be seen, that indeed the transmission of the LCP or reflection of RCP waves are nearly equal to their values of 50% in the absence of absorption, however, reflection of LCP and transmission of RCP waves are equally reduced by almost a factor of two. This demonstrates that LCP reflection as well as RCP transmission is the result of coupling to the long-lived localized state.

In order to understand the nature of the defect mode, we examine its polarization by decomposing the wave inside the medium into two circular polarized components for incident LCP (Fig. 4b) and RCP (Fig. 4c) waves in a sample with thickness $L = 160P \approx 4L_{co}$. For each polarization of the incident wave, the components propagating in the forward direction are shown.

For both senses of circular polarization of the exciting radiation, at any sample thickness, the wave near the defect is RCP and the wave intensities for the two directions of propagation are equal. Thus, the defect mode is a standing RCP wave near the defect, where most of the energy is stored. Closer to the boundaries, however, the wave is in general elliptically polarized and determines the polarization of the emission emerging from the structure at the defect frequency. An incident wave with the same polarization is most efficiently coupled to the localized state. The orthogonally polarized wave is completely decoupled from the localized state.

Let us consider the coupled waves at various values of $L/L_{co}$. For $L \ll L_{co}$, the incoming mode that excites the localized mode is a circularly polarized wave with the same handedness as the structure, RCP in our case. The orthogonal wave, LCP, cannot excite the localized state. At $L$ =

$L_{co}$, the localized mode is most efficiently excited by a linearly polarized wave, whereas the perpendicularly polarized wave does not excite the localized state. For $L >> L_{co}$, the wave that couples to the localized state has circular polarization opposite to the handedness of the structure (LCP in our case), while the orthogonal wave, (RCP) does not excite the localized mode. It is interesting to note, that for $L = L_{co}$, the incoming linearly polarized wave is split 50/50 in reflection/transmission. The polarization, which is coupled to the localized state, is transmitted as RCP and reflected as LCP, while the reverse holds for the perpendicularly polarized incident wave.

As mentioned above, both the transmission of the RCP wave and reflection of LCP wave can occur only via the excitation of the localized state. The transmission of RCP is therefore the coupling coefficient between the RCP wave and the localized state. The same is true for LCP radiation. The sum of the corresponding coupling coefficients for the energy is therefore equal to unity for any $L$, as is seen to be unity in Fig. 3 for the sum of the transmission of RCP and reflection of LCP.

In conclusion, we have shown that the chiral twist defect creates a single circularly polarized localized mode in contrast to a defect in a binary layered medium, which produces two orthogonally polarized degenerate localized modes. As a result, right or left circularly polarized incoming waves can be decomposed into two orthogonal components one of which is coupled to and another is decoupled from the chiral twist defect mode. The variation with thickness in the amplitudes of these components results in a crossover in reflection and transmission of circularly polarized waves. As a result, the intensity at the site of the localized mode saturates instead of scaling exponentially. Nonetheless, the intensity enhancement can be greater than $10^3$ as in seen in Fig.4b. This gives a resonant Q-factor of more than 30,000. Much larger values of Q are calculated for structures with lower anisotropy. Achieving such Q-factors in anisotropic media such as polymeric CLCs or sculptured thin films would open the way for a variety of active and passive integrated optical devices.

**Acknowledgements**

The support of the Army Research Office (DAAD190010362) and by the National Science Foundation is gratefully acknowledged.

**Figure Captions**

Fig. 1. Schematic of the chiral twist defect produced in periodic cholesteric structure with pitch $P$ and period $a$ by rotating top portion. Unmodified structure is shown on the left.

Fig. 2. Transmittance of right and left circularly polarized light versus wavelength for different structure thickness. (a) Structure thickness is less than crossover thickness (~ 0.45 $L_{co}$) and (b) structure thickness is bigger than crossover thickness (~ 1.8 $L_{co}$).

Fig. 3. Transmittance of right and reflectance of left circularly polarized light versus structure thickness $L$ at the resonance wavelength. $P$ is the structure pitch.

Fig. 4. Energy density inside the structure of pitch $P$ versus the coordinate $z$ along the direction of wave propagation. (a) Energy density of the forward propagating wave for different polarizations of the incident wave and sample thickness. (b) and (c) Decomposition of the forward propagating wave into right and left circularly polarized components for left (b) and right (c) circularly polarized incident wave.

Fig. 5. Inverse relative linewidth versus relative sample thickness $L/P$.

Fig. 6. RCP and LCP transmittance and reflectance at crossover length in presence of absorption.

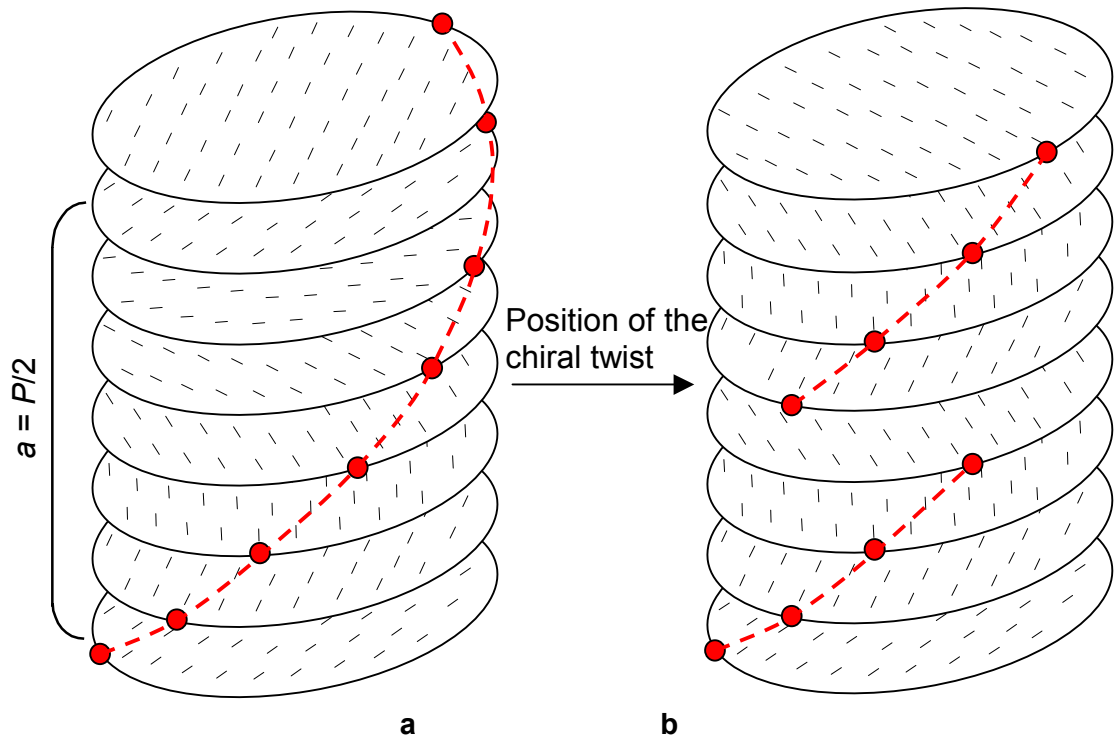

Fig. 1.

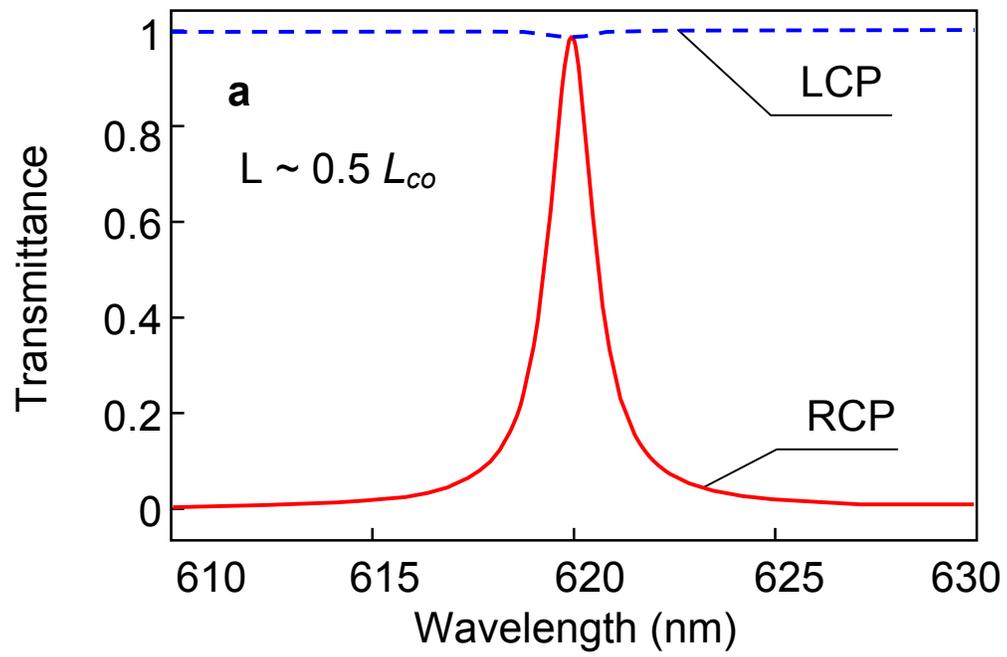

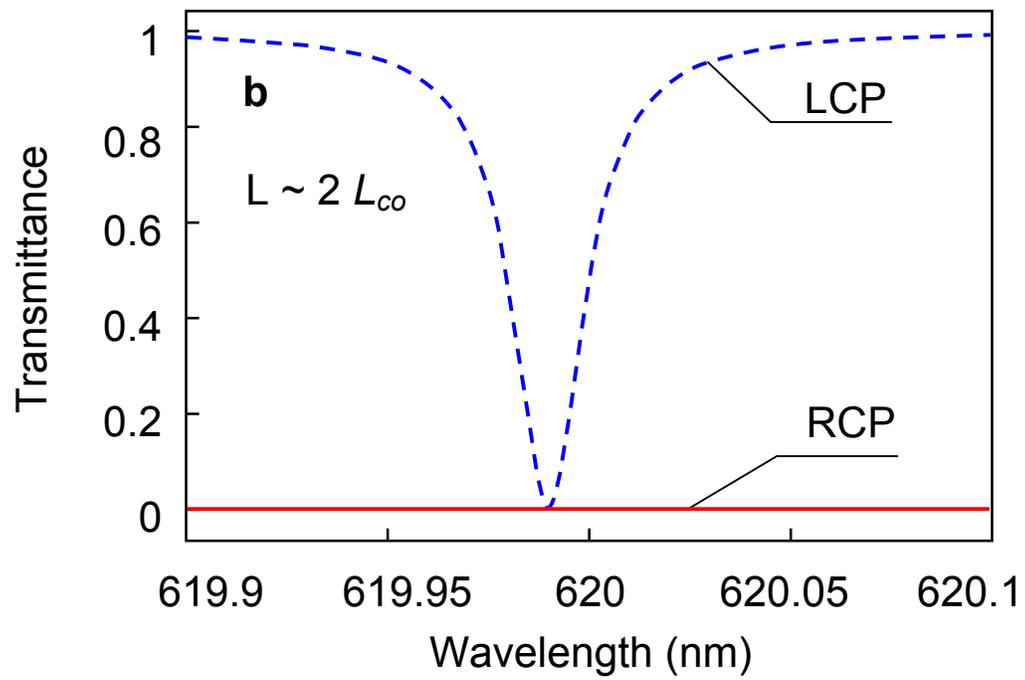

Fig. 2

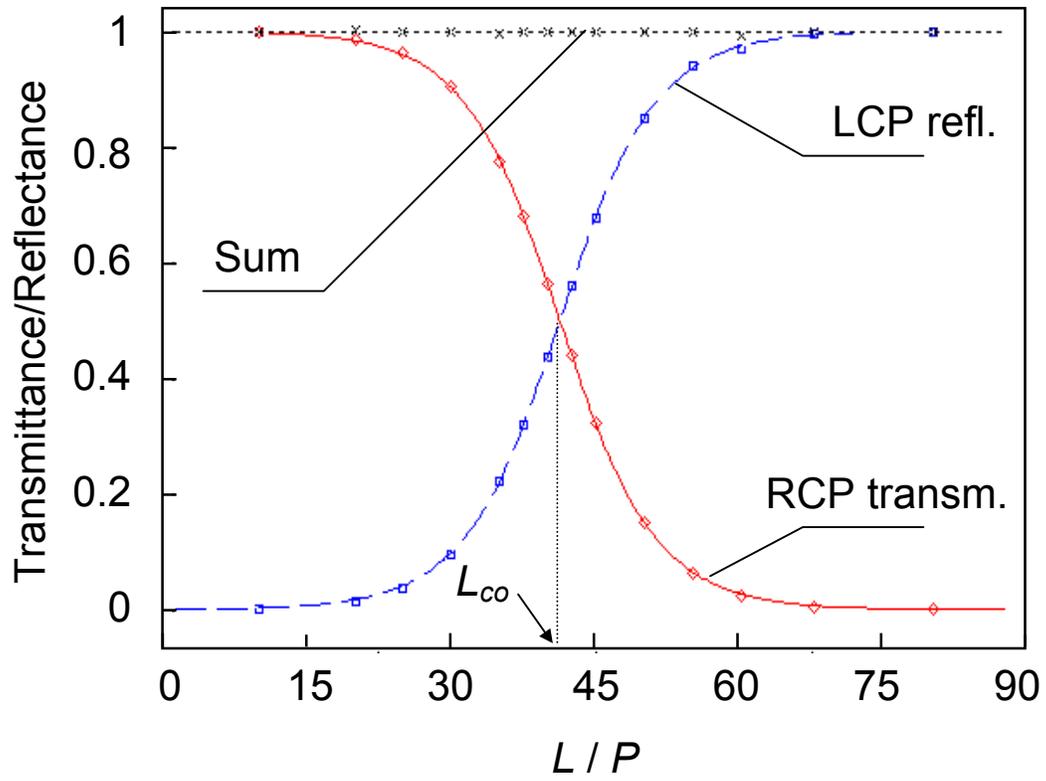

Fig. 3

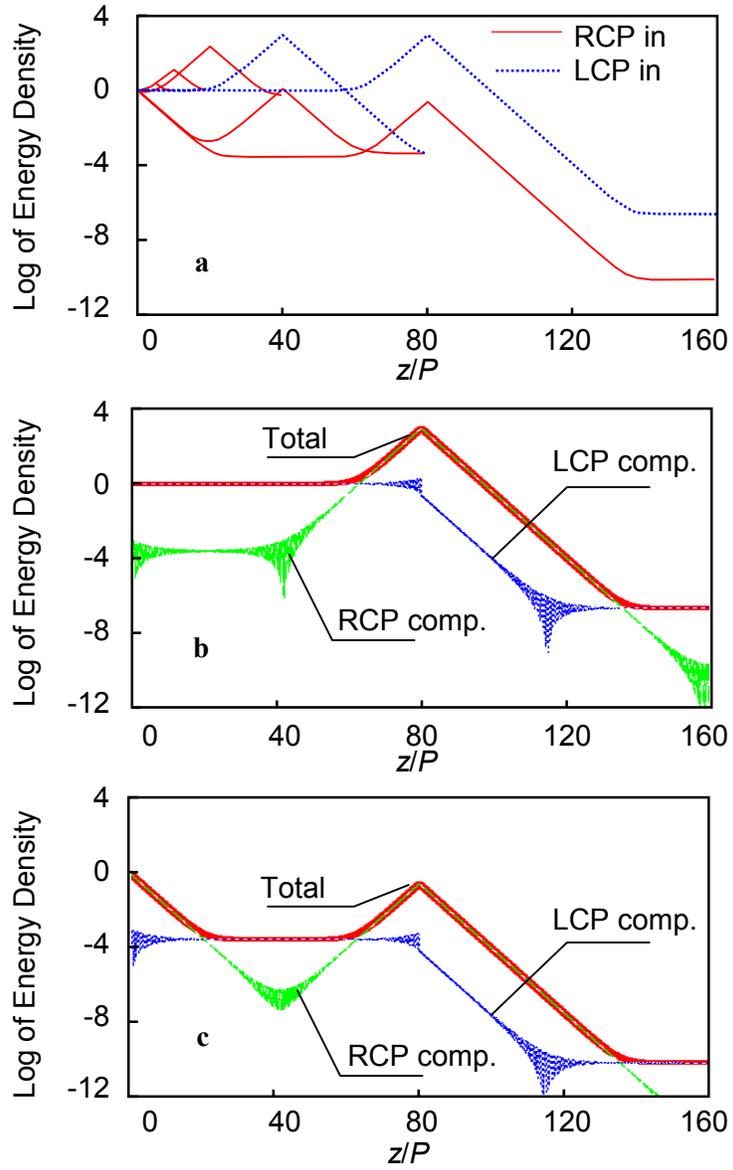

Fig. 4

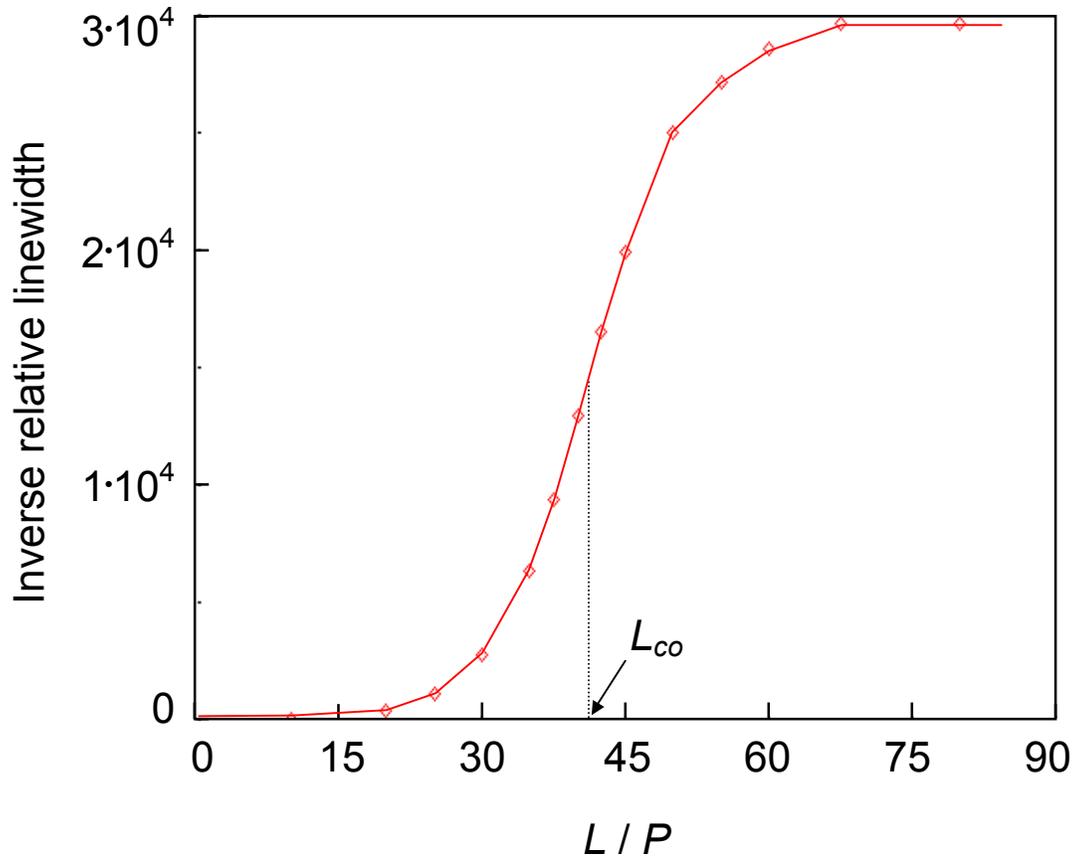

Fig. 5

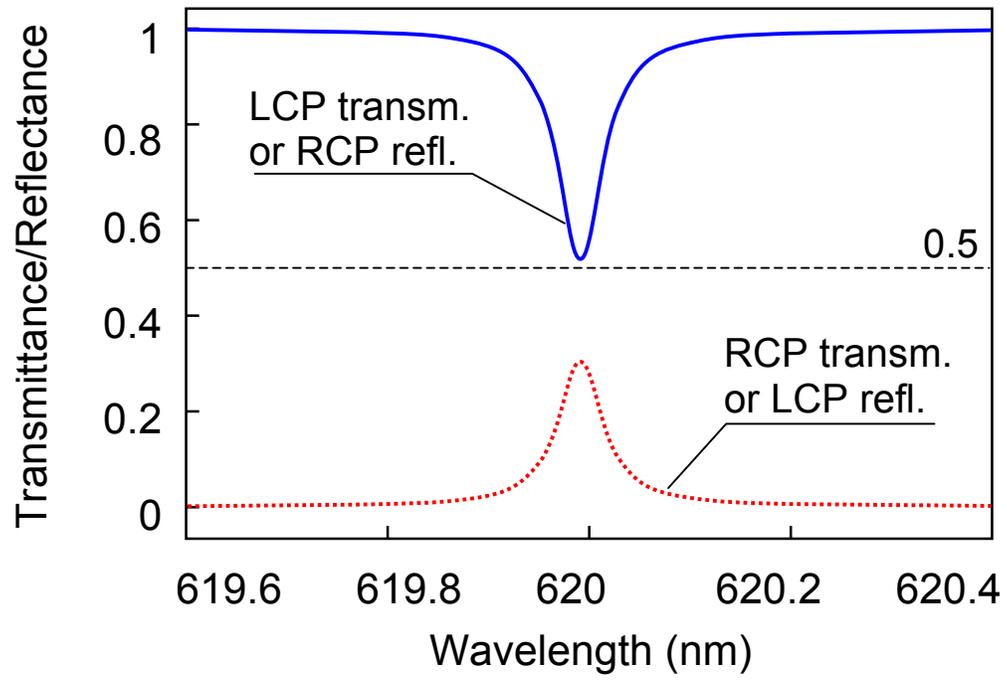

Fig. 6